\documentclass[twocolumn,showpacs,amsmath,amssymb,prd,nofootinbib]{revtex4}
\usepackage{epsfig}
\def\sss{\scriptscriptstyle}
\def\_#1{_{\sss #1}}
\def\!#1{^{\sss #1}}

\def\beq{\begin{equation}}
\def\eeqno#1{\label{#1}\end{equation}}

\def\az{a_{\scriptscriptstyle 0}}
\def\A{\mathcal{A}}

\def\azg{\A_0}
\def\baz{\bar a_{\scriptscriptstyle 0}}
\def\lz{\ell_{{\scriptscriptstyle 0}}}

\def\rar{\rightarrow}
\def\s{\sigma}
\def\l{\lambda}

\def\Lam{\Lambda}

\def\f{\phi}
\def\grad{\vec\nabla}

\def\gf{\grad\phi}

\def\o{\omega}

\def\L{\mathcal{L}}

\def\gh{g^{1/2}}

\def\gmn{g_{\m\n}}

\def\m{\mu}
\def\a{\alpha}
\def\b{\beta}

\def\d{\delta}

\def\z{\zeta}
\def\t{\theta}

\def\n{\nu}

\def\F{\mathcal{F}}

\def\D{\Delta}

\def\lM{\ell\_M}
\def\lU{\ell\_U}
\def\lH{\ell\_H}
\def\dsr{\ell\_\Lam}
\def\alam{a\_\Lam}
\def\la{\ell_a}
\def\aU{a\_U}
\def\MM{M\_M}
\def\MU{M\_U}
\def\rM{r\_M}
\def\fm{\varepsilon}
\begin{document}
\title{The $\az$ - cosmology connection in MOND\footnote{Based on a talk at `BonnGravity2019 -- The functioning of galaxies: challenges for Newtonian and Milgromian dynamics', Bonn, September 2019.}} \author{Mordehai Milgrom }
\affiliation{Department of Particle Physics and Astrophysics, Weizmann Institute}

\begin{abstract}
I limelight and review a potentially crucial aspect of MOND: The near equality of the MOND acceleration constant, $\az$ -- as deduced from local, galactic phenomena -- and cosmological parameters. To wit, $\az\sim c H_0\sim c^2\Lam^{1/2}\sim c^2/\lU$,
where $H_0$ is the present value of the Hubble-Lema\^{i}tre constant, $\Lam$ is the `cosmological constant', and $\lU$ is a cosmological characteristic length; e.g., the Hubble distance, or the de Sitter radius associated with $\Lam$.
In itself, this near equality has some important phenomenological consequences, such as the impossibility of black holes, and of cosmological strong lensing, in the MOND regime.
More importantly perhaps, this `coincidence' may be a pointer to the `FUNDAMOND' -- the more basic theory underlying MOND phenomenology.
The manners in which such a relation emerges in existing, underlying scheme of MOND are also reviewed, interlaced with examples of similar relations in other physical systems, between apparently-fundamental velocity, length, and acceleration constants. Such analogies may point the way to explanation of the MOND `coincidence'.

\end{abstract}
\pacs{04.50.Kd, 95.35.+d}
\maketitle

\section{Introduction}
The acceleration constant, $\az$, is central to MOND dynamics \cite{milgrom83a,milgrom83b,milgrom83c} (reviewed in Refs. \cite{fm12,milgrom14c}).
In several MOND laws it appears in the role of a `boundary constant', separating the validity domain of standard dynamics for accelerations much above $\az$, and the scale-invariant, deep-MOND dynamics below $\az$.
It also appears ubiquitously in MOND laws that pertain to the deep-MOND regime itself, where scale invariance dictates that $\az$ appears only in the product $\azg=G\az$. (See \cite{milgrom14a} for a detailed discussion of `MOND laws' and how they follow from the basic tenets of MOND.) Such roles of $\az$ are similar to those of $\hbar$ in quantum physics, or of $c$ in relativity. All three play the roles of boundary constants; they do not appear in phenomena in the unmodified regime -- the classical, pre-1900 physics; and they each appear ubiquitously in phenomena in the modified regime.
\par
It has been realized right from the outset \cite{milgrom83a} that the value of $\az$ is near in value to cosmologically significant acceleration parameters. In Ref. \cite{milgrom83a} it was noted and commented on that $\az\sim cH_0$, with $H_0$ the present expansion rate of the Universe -- the Hubble-Lema\^{i}tre constant. Then, in Ref. \cite{milgrom89} the connection was also made with a hypothetical cosmological constant, $\Lam$, (on which only upper limits existed at the time). Then, with more relevance \cite{milgrom94}, after the first hints of the presence of a cosmological constant emerged \cite{efstathiou90}. And then \cite{milgrom99}, with final concreteness, after a cosmological-constant-like effect in cosmology was definitely identified. Reference \cite{milgrom99} also offered the first proposal for accounting for the coincidence (as discussed below).
\par
The value of $\az$ as best known today is (e.g., Ref. \cite{li18})
\beq \az\approx 1.2\times 10^{-8}~{\rm cm~s^{-2}}, \eeqno{azero}
to within 10-20\%.
The many determinations of $\az$ to date, starting with the first MOND analysis of rotation curves (by Steve Kent, reported in Ref. \cite{milgrom88}) yielded values in this range.
\par
With $c$ at our disposal, we can construct acceleration parameters of cosmological bearing from $H_0$ (of dimensions time$^{-1}$), and from $\Lam$ (of dimensions length$^{-2}$)
\beq  a\_H\equiv cH_0,~~~~~~\alam\equiv c^2(\Lambda/3)^{1/2}  \eeqno{acca}
($\Lam/3$ is the natural way $\Lam$ enters), which I shall refer to jointly as $\aU$.
With the presently determined values of $H_0$ and $\Lam$,
the $\az$-cosmology near-equalities are
\beq \baz\equiv 2\pi\az\approx a\_H\approx \alam.  \eeqno{coinc}
 The fact that $\alam$ -- supposedly a constant -- and $a\_H$ -- which uses the present day value of the variable expansion rate -- are roughly the same today, is a cause for pondering in itself.
\par
Relations (\ref{coinc}) can be written in other, possibly-revealing, forms in terms of the `MOND length' and `MOND mass',
\beq \lM\equiv c^2/\az, ~~~~\MM\equiv c^4/\azg: \eeqno{mlength}
as
\beq \lM\approx 2\pi\lU,~~~~~~~~~\MM\approx \MU  , \eeqno{elrel}
where $\lU$ stands for either the Hubble radius $\lH\equiv cH_0^{-1}$, or the de Sitter radius associated with $\Lam$, $\dsr\equiv (\Lam/3)^{-1/2}$, and $\MU$ is total mass within the Hubble radius, or the de Sitter radius.
\par
The mere numerical near equality in eq.(\ref{elrel}), even without delving into its significance, has some important implications for phenomenology.
\par
But far beyond this, as discussed in many previous accounts of the $\az$-cosmology relation, it may hint at a deeper stratum underlying MOND phenomenology. If indeed not just a coincidence, this relation implies almost inescapably, that MOND, as we now describe it, is an effective, approximate theory. The connection may also imply that trying to understand cosmology within local MOND -- as general relativity (GR) cosmology is described within GR -- is not the right path. Rather, that we will eventually understand MOND as a part and parcel of cosmology (see also Sec. \ref{rbh}).
\par
We can exemplify this expectation, and perhaps dispel some of the mystery in the coincidence, by pointing to analogous relations, in other physical systems, between apparently-fundamental constants of dimensions length, velocity, and acceleration.
\par
One familiar such example relates the `constant', near-Earth-surface, free-fall acceleration, $g\_G$, the Earth radius, $R\_\oplus$, and the escape speed from the Earth's surface, $V_e$:
\beq g\_G=V^2_e/2R\_\oplus.\eeqno{earth}
At some effective (approximate) level that applies when we limit ourselves to Earth-surface experiments,\footnote{I.e., if we knew nothing about the Universe beyond Earth, but we could throw things upwards and notice that when thrown with seeds exceeding $V_e$ they do not return.} the three parameters may seem as unrelated fundamental constants that satisfy a mysterious numerical relation. It is only through Newtonian gravity, relating $g\_G$ and $V_e$ to the mass and radius of the earth, that this relation is understood as a corollary.
And, we could not have deduced Newtonian gravity in a framework in which $g\_G$ is a fundamental constant, with only near-Earth-surface experience, and without exploring objects far above it. The lesson for MOND is evident.
\par
In Sec. \ref{practical}, I list some immediate phenomenological corollaries of the numerical proximity of $\lM$ and $\lU$.
In Sec. \ref{why}, I review briefly some of the ways in which the $\az$-cosmology connection emerges in existing MOND theories or underlying schemes. I also discuss there some analogue relations in more familiar physical systems. Section \ref{summary}
is a summary.
\section{$\lM\sim\lU$: Phenomenological implications}
\label{practical}
There are many local (e.g., galactic) phenomena in which $\az$ enters as an acceleration constant. But inasmuch as these phenomena are nonrelativistic, the $\az$-cosmology connection does not come into play in them: This connection -- which involves $\az$, and the cosmological $\lU$, but also $c$, can come to bear explicitly in the phenomenology only if the speed of light also enters. This is the case in strong-gravity phenomena (such as black holes and cosmology), where the gravitational potential is $\f \approx c^2$. It is also the case in phenomena involving objects moving at speeds near $c$, such as electromagnetic or gravitational waves, or highly relativistic massive particles, such as high-energy cosmic rays.
\par
In the Earth-gravity analog, as long as we are dealing with phenomena at speeds much below the escape speed, the `coincidence' $g\_G=V^2_e/2R\_\oplus$ is not part of the observed phenomenology; only $g\_G$ enters. Only when we deal with high-speed phenomena -- such as a satellite in near-Earth orbit with $V\approx V_e/\sqrt{2}$ -- does the `coincidence' enter explicitly.

\subsection{No deep-MOND black holes on sub-Universe scales\label{rbh}}
A system -- such as the near vicinity of a black hole -- is subject to `strong gravity' if its characteristic metric departure from flat Minkowski, $\d g \not\ll 1$. For example, if we write the $00$ component of the metric as $-1-2\f/c^2$, where $\f$ becomes the Newtonian gravitational potential in the nonrelativistic limit, then such systems are characterized by $\f\sim c^2$.
On the other hand, for a system to be in the MOND regime, the gravitational accelerations in it should be smaller than $\az$. This means that the metric connections, or the derivatives of the metric, should be $\lesssim\lM^{-1}$.\footnote{Of course, at any point we can take coordinates where the metric derivatives vanish; but this cannot be done everywhere in the system, and I am speaking here of characteristic values.}
\par
If $R$ is the characteristic size of the system, then requiring both strong gravity and low accelerations, imply $R\gtrsim c^2/\az=\lM$. But, since $\lM\approx 2\pi\lU$, this means that the system has to be larger than even the present observable Universe, and also larger than the de Sitter radius associated with the cosmological constant.
\par
Taken at face value, this result tell us that there is no phenomenological need for a deep-MOND, `strong gravity' relativistic theory, as phenomena in this domain cannot be probed.
I view this result also as an indication -- alluded to above -- that the relativistic framework for MOND will have to be understood as a corollary of cosmology, and not as a prerequisite theory within which cosmology will be derived.

\subsection{No deep-MOND, `cosmological' strong lensing}
The MOND radius of a mass $M$,
\beq \rM\equiv \left(\frac{MG}{\az}\right)^{1/2}, \eeqno{mondradius}
has a role analogous to that of Schwarzschild radius in relativity: If the mass is well within its MOND radius, then $\rM$ marks the transition from standard dynamics at $r\ll\rM$ to the MOND regime at $r\gtrsim\rM$.
\par
The Einstein radius, $r\_E$, of a gravitational lens of mass $M$, around and within which strong lensing occurs if the mass itself is well within $r\_E$, is given by
\beq r\_E=\left(\frac{4GM}{c^2}\frac{D\_{ls}D\_{l}}{D\_{s}}\right)^{1/2},  \eeqno{eins}
where $D\_{l},~D\_{s},~D\_{ls}$ are, respectively, the distances to the lens, to the source, and the distance from the lens to the source.
We thus have
\beq \frac{r\_{E}}{r\_{M}}=\left(\frac{4D\_{ls}D\_{l}}{D\_{s}\lM}\right)^{1/2}\approx \left(\frac{4D\_{ l}}{\lM}\right)^{1/2}, \eeqno{einmond}
where in the second, near equality, I assumed that the source is rather farther than the lens so $D\_{s}\approx D\_{ls}$.
\par
Thus, because $\lM\approx 2\pi\lU$, the Einstein radius is within the MOND radius, and under these circumstances strong lensing cannot occur in the deep-MOND regime.

\subsection{Gravitational acceleration in the presence of gravitational waves produced by black holes is $\gg\az$ even after traveling cosmological distances}
Very near `strong gravity' sources of gravitational waves,\footnote{Such as the near vicinity to, and just prior to the merger of two black holes of comparable mass.} the wave field itself is of `strong gravity' (in the relativistic sense).
This field is also of very high accelerations in the context of MOND: The `natural' value of the acceleration in the wave in this `strong gravity' regime is $g\_W\sim c^2/\l$ ($\l$ is the wavelength, which is of the order of the Schwarzschild radius of the radiating masses). So, $g\_W/\az\sim \lM/\l\gg1$ .
This means that in GR-compatible MOND theories -- namely, theories that tend to GR in the limit of high accelerations ($\az\rar 0$) -- the emission process of such `strong-gravity' systems should be describe by GR to a high accuracy.
\par
The wave field becomes of `weak gravity' once it travels only several gravitational radii away from the source. However, the wave field can remain of high accelerations (i.e., with $g\_W\gg \az$) even after traveling for cosmological distances (see more details in Ref. \cite{milgrom14b}).
\par
The gravitational-wave strain produced at a distance $D$ from a radiating system of total mass M can be written as
\beq  h= f\frac{MG}{Dc^2},   \eeqno{gw}
where $f$ is a dimensionless constant that depends on various dimensionless parameters of the radiating system, such as the distribution of the mass in the system (e.g., the mass ratio in a binary), the velocities in the system (relative to $c$), geometrical characteristics, etc.
For nearly merging black holes with similar masses $f$ is of order 1.
\par
The gravitational acceleration in the wave can be estimated as $g\_W\approx c^2h/(\l/4)$, which gives $g\_W/\az\approx 4h\lM/\l$. Write $\l=qR_s$, where $R_s=2MG/c^2$ is the Schwarzschild radius of the radiating system, then
\beq \frac{g\_W}{\az}\approx \frac{2f}{q}\frac{\lM}{D}.  \eeqno{gita}
So, for a given dimensionless ratio $f/q$ (which can be small, but for highly-efficient, relativistic emitters can be larger than unity), what sets the distance over which $g\_W$ decays below $\az$ is the MOND length $\lM\approx 2\pi\lU$.

\subsection{Gravitational-Cherenkov life times of high-energy cosmic rays}
In some relativistic theories that depart from GR,
there are gravitational-wave modes that propagate with velocities $V\_{GW}<c$.
In particular, this may be the case with some relativistic formulations of MOND.
The question then arises whether very-high-energy cosmic rays, with velocities $V\_{CR}>V\_{GW}$, would emit the subluminal modes by Cherenkov radiation at high enough rates to stop them from propagating long distances. In the context of MOND, the issue was considered in Ref. \cite{milgrom11}.
\par
Succinctly put, the arguments are as follow:
The energy radiated per unit time per unit
wave vector, $k$, in gravitational Cherenkov radiation, by a relativistic particle of momentum $p$ is
 \beq \frac{d^2E}{dk dt}= \frac{SGp^2k}{c}. \eeqno{spec}
Here, $S$ is a dimensionless factor that depends on the specific theory (through, e.g., $V\_{GW}/c$), on $\hbar k/p$, and on the particle's velocity $\b=v/c$. The
total energy-loss rate, $dE/dt$, is gotten by integrating expression (\ref{spec}) over $k$ and is dominated by the contribution of the high wave numbers dictated by some, possibly gradual, cutoff $k_c$.
\par
The resulting expression for the loss distance is
 \beq D^{-1}_{loss}\equiv E^{-1}\frac{dE}{dx}=
 \frac{Gpk_c^2}{c^3}Q,
   \eeqno{rate}
where $Q$ is a dimensionless quantity that depends on dimensionless
parameters of the theory, on $\b$, on $V\_{GW}/c$, and on the internal structure of the
particle.
\par
It was argued in Ref. \cite{milgrom11} that for MOND theories that are compatible with GR, the cutoff $k_c$ is at most of order of $\rM^{-1}$, where $\rM$ is the MOND radius of the particle given from eq. (\ref{mondradius}) by
 \beq r\_M=(Gp/c\az)^{1/2}\approx 3\times 10^{-12}
 (cp/{\rm Gev})^{1/2}{\rm cm} \eeqno{jupa}
($pc$ is the energy, or mass, of the particle in the laboratory frame).
\par
This conclusion was based on the argument that within $\rM$, GR prevails to a high accuracy, and the subluminal modes that are foreign to GR cannot propagate there, and so are not emitted at all if their wavelength is $\lesssim \rM$.
Effectively, the particle carries around it a bubble of radius
$r\_M$, devoid of the subluminal waves. This conclusion was further buttressed in Ref. \cite{milgrom11} by reference to explicit calculations of the electromagnetic Cherenkov radiation from a point charge moving with velocity $V$ along the axis of a cylindrical vacuum cavity of radius $R$ in a dielectric medium in which the speed of light is $<V$. The Cherenkov spectrum is indeed found to be cut off (exponentially) at wavenumbers larger than $R^{-1}$, for the same reason adduced above for the gravitational case.\footnote{Explicit calculations for the even less restricted case of a point charge moving in a vacuum half-space parallel to, and at a distance $R$ from, the surface of a dielectric half space, also show that he spectrum is cut off exponentially for $k>R^{-1}$.}
\par
Putting all this together, the `coincidence' (\ref{coinc}) tell us that the distance over which the particle losses a large fraction of its energy is given by
\beq D_{loss}=q\ell\_M\approx 2\pi q \lU,  \eeqno{ratebh}
where the dimensionless prefactor $q$ has
a similar role to that of $Q^{-1}$ in Eq.(\ref{rate}). It depends on
parameters of the theory and on the wave and particle velocities (in
units of $c$). It insures, for example, that
$D_{loss}$ is infinite if the particle speed is below the wave
speed.
\par
We see then that, quite generally, due to the coincidence (\ref{coinc}), in GR-compatible MOND theories, Cherenkov radiation of subluminal gravitational modes does not stop relativistic particles from moving over cosmological distances. Reference \cite{cl17} has fallaciously contested this conclusion.\footnote{They do so by claiming to refute an argument that they attribute to me, but which was not the argument I used. According to Ref. \cite{cl17}, my argument was that due to MOND effects the particle radiates as if it acted as an extended particle of size $\rM$, and thus due to incoherent destructive interference by radiation from different parts, radiation with wavelength shorter than $\rM$ is cut off. They then go on to question this argument. However, this has nothing to do with my actual argument. As explained in the main text, I treated the emitter as a point mass, and no coherence requirement was recoursed to. I argued, instead, that within $\rM$, GR rules, and so no Cherenkov modes can exist there. It is this that cuts off modes with wavelengths shorter than $\rM$. (Reference \cite{cl17} may have been red-herringed by my mentioning in passing that explicit calculations of Cherenkov radiation by extensive bodies also show a spectral cutoff at wavelength beyond the emitter's size.)}

\section{Why is $\lM\sim\lU$? \label{why}}
The interesting phenomenological ramifications discussed in Sec. \ref{practical}
are simple consequences of the near {\it numerical} proximity $\lM\approx 2\pi\dsr$, irrespective of its origin or significance.
But, shall we take it seriously as an important clue to the physics underlying MOND phenomenology? Shall we spend time in trying to explain it? Shall we require from a candidate MOND theory that it account for this relation?

History has taught us that it sometimes behooves us to mind our coincidences, lest we miss a chance for an important discovery. In a famous mid-19th-century example, Weber and Kohlrausch noticed that their measured value of the so-called `ratio of electromagnetic and electrostatic units' was near in value to the then known speed of propagation of light, but seem not to have made much of it, leaving it to Maxwell to conclude, some years later, that light is an electromagnetic wave!
\par
Historical precedents aside, deciding between a promising clue and a red herring remains somewhat a matter of personal choice and taste.
In the present instance, I feel that this coincidence will turn out to be an important lead. This belief is based on the following:  1. the phenomena that hinge on the constants involved in the coincidence -- namely, galaxy dynamics and cosmology -- are both governed by gravity, and similar anomalies, which are otherwise attributed to dark matter, appear in both. 2. There are examples of similar velocity-size-acceleration-constants relations in other physical systems that would appear as mere numerical coincidences if the underlying physics where not known (see below).
3. As discussed below, there are already a number of possible explanations of such a relation, making it, in the least, plausible that the relation does have some underlying physical reason.
\par
If the coincidence does have underlying physical reasons, then it points to MOND phenomenology being emergent from a deeper `FUNDAMOND' as in the above example of $g\_G$.
\par
The question whether $\az$ is to be tied to $H_0$ or to $\Lam$ (or to both in some way) is still moot.
\par
In what follows I describe some ways in which a relation such as (\ref{coinc}) arises in various underlying MOND schemes.
\subsection{Naturalness, or parsimony in introducing new constants}
By `naturalness' of a physical theory we usually mean that if apparently-fundamental constants of the same dimensions appear in the theory, then it is natural to expect that their ratio is of order unity. Alternatively put, it is `unnatural' for a theory to involve dimensionless constants that differ much from unity (unless there are good reasons to be provided for this).
\par
If `unnaturalness' in this regard appears in a theory, it is many times taken as a sign that the theory is only an approximation of a deeper, `natural' theory, and that we should seek explanations for the plurality of same-dimensions constants. This precept has, for example, driven much research in explaining the `unnaturalness' appearing in the `standard model of particle physics'.
\par
But clearly, this precept is not carved in stone, and `unnaturalness' may appear naturally in a physical system in ways that do not tell on a deeper, significant underlying theory.
\par
Be that as it may, in some MOND theories requiring such naturalness does imply the coincidence (\ref{coinc}).
\par
This happens as follows:
In some effective-field, relativistic MOND formulations, the modification is effected, schematically speaking, by adding to the Einstein-Hilbert Lagrangian density of GR, $\L\_{EH}\propto (c^4/G)\gh R$, ($R$ is the Riemannian curvature scalar)  a gravitational Lagrangian of the form
\beq (c^4/G)\gh\ell^{-2}\F(\ell^2 Q).  \eeqno{repla}
This is the case in MOND adaptations of Einstein-Aether theories \cite{zlosnik07}, in BIMOND (bimetric MOND) \cite{milgrom09b}, in the bimetric massive gravity of Ref. \cite{blh17}, and in the noncovariant theory described in Ref. \cite{milgrom19}.
Here, $\F$ is a dimensionless function, $\ell$ in front of it is a length constant that gives the Lagrangian the correct dimensions. $Q$ is constructed from the first (space-time) derivatives of the gravitational degrees of freedom and has dimensions of length$^{-2}$. For example, in MOND adaptations of Einstein-Aether theories, $Q$ is a scalar, quadratic expression in derivatives of the vector, `aether' field; in BIMOND, $Q$ is a scalar, quadratic expression in the difference of the Levi-Civita connections of the two metrics; and in the theory described in Ref. \cite{milgrom19}, $Q$ is a non-scalar, quadratic in the Levi-Civita connection of the metric.
\par
Because we take the same constant $\ell$ to appear inside and outside $\F(x)$ (where it plays different roles), naturalness here means that $\F(x)$ itself does not involve dimensionless constants that much differ from unity. ($\F$ can still take small or large values when its dimensionless argument takes values that much differ from unity.)
\par
To get MOND in the nonrelativistic limit, where
$\L\_{EH}\propto G^{-1}(\gf)^2$ (up to immaterial derivatives, such as $\D\f/c^2$) we need $Q\rar (\gf)^2/c^4$, where $\f$ is the nonrelativistic gravitational potential (or the main contribution to it in the MOND regime).\footnote{$F(R)$ theories are gotten by taking $Q=R$, but these do not have the appropriate nonrelativistic limit to give MOND (see Ref. \cite{milgrom15}).}
The argument of $\F$ becomes in this limit $\ell^2 (\gf)^2/c^4\equiv (\gf)^2/\az^2$,
where we identify $c^2/\ell\equiv\az$. Thus $\ell$ is now identified as the MOND length $\lM$, and the nonrelativistic modifying Lagrangian becomes
\beq \bar\L\_M=G^{-1}\az^2\F[(\gf)^2/\az^2].  \eeqno{nrlag}
\par
On the other hand, remembering that a cosmological constant appears in the GR Lagrangian as
$(c^4/G)\gh\Lam$, we see from expression (\ref{repla}) that any constant contribution to $\F(x)$, which, naturalness dictates, is of order unity, gives a cosmological constant
$\Lam\sim \ell^{-2}=\lM^{-2}$, thus yielding the `coincidence' (\ref{coinc}).
\par
Note that the above naturalness argument is not the same as adding to the Lagrangian a cosmological-constant term, $(c^4/G)\gh\Lam$,
and then invoking naturalness to deduce that $\Lam$ is of the same order as $(\az/c^2)^2$.
In our theories, the length constant, $\ell$, must appear, for dimensional reasons, in the two places in the same Lagrangian term, and so they both come from the same underlying physics. Thus, $\ell$ is not added `by hand' as a separate cosmological-constant term, which might come from different underlying physics.
\par
Finally, note that a (different) requirement of naturalness enters already at the basic level of MOND, without even reference to the $\lM-\lU$ coincidence: One of the basic axioms of MOND is that the Newtonian-to-MOND transition occurs around $\az$ (the role of $\az$ as a `boundary constant'). But, for the basic tenets to yield the `MOND laws', it is also assumed in small print that the transition occurs over an acceleration range that does not much exceed $\az$ (see, e.g.,  Ref. \cite{milgrom14a}). Otherwise, MOND would involve two, very different, acceleration constants, and some of the MOND laws would not have been valid.
\par
In the same vein, we take it for granted that there are no two $\hbar$s in quantum theory, or two $c$s in relativity.

\subsection{Emergent dynamics in `boxed' systems}
There are many instances of physical systems that are boxed in a volume, or that are in the presence of boundaries, of characteristic length $\ell$, and whose internal dynamics know about $\ell$ in one way or another.
\par
This can happen, for example, because the `box' dictates certain boundary conditions directly on degrees of freedom of the system under consideration, or because the presence of the `box' affects `hidden' fields that, in turn, interact with the system, but that are eliminated in the emergent dynamics for our sub system.\footnote{For example, the geometry of the background space-time may affect its quantum vacuum, as in the Gibbons-Hawking effect, in a way that knows about the background geometry, or a box with conducting walls affects the electromagnetic vacuum in it, as in the Casimir effect.}
\par
If there are other relevant `fundamental' dimensionful constants besides $\ell$, then $\ell$ may enter the dynamics not (only) as a length, but also in a combination with that constant. By `enter as a length' I mean that it appears in the effective dynamics of the system as a constant with dimensions of length. `Enter as acceleration' describes what happens in MOND, which revolves on the effective constant $c^2/\ell$.

For example, if there is a fundamental velocity, $V_f$, then a combination of $\ell$ and $V_f$, may enter as a `fundamental constant' in the effective theory of our system, such as an acceleration $a_f\equiv V^2_f/\ell$, or a characteristic time $t_f\equiv\ell/V_f$.
\par
A familiar example is a quantum mechanical system; e.g., of otherwise-free electrons, in a box of size $\ell$. Here, besides $\ell$, we have a fundamental action, $\hbar$. So, $\ell$ may enter the dynamics not only as a length, but also as a `fundamental' momentum $P_f\equiv \hbar/\ell$, with particle dynamics being classical in the limit $P\gg P_f$, and manifestly quantum in the opposite case.
\par
\begin{figure}[h]
	\centering
\includegraphics[width=0.6\columnwidth]{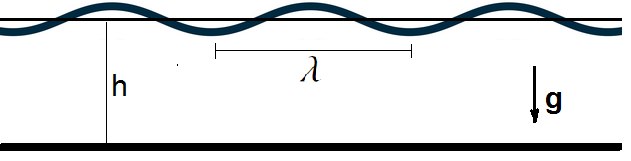}
\caption{Gravity waves -- parameter definition.}		\label{waves}
\end{figure}

Another edifying example is that of gravity waves in a reservoir of an ideal fluid of constant depth $h$ (playing the role of the `box' size) in a gravitational field of constant acceleration $g$ -- the second effective constant (see Fig. \ref{waves}).
The dispersion relation for surface waves of small amplitude (of amplitude $\d h\ll h$) is
\beq \o^2=gk\cdot{\rm tanh}(kh), \eeqno{disper}
where $k\equiv 2\pi/\l$ is the wave number. The long-wavelength ($kh\ll 1$) phase (and group) velocity is $c^2\_\ell\equiv gh$.
\par
This two-dimensional description of the system is clearly only an approximate, effective account of the full three-dimensional hydrodynamical behavior. It requires, e.g., that bulk waves communicate with the bottom on time scales much shorter than $\o^{-1}$ , or that the bulk speed of sound is  $c^2_b\gg c^2\_\ell hk=h^2 gk$.
\par
Rewrite eq. (\ref{disper}) in terms of the constants $g$ and
\beq M\bar G\equiv gh^2=c^2\_\ell h  \eeqno{consta}
(instead of $h$), and in terms of the degrees of freedom `radius'
\beq r_\o\equiv (gh)^{1/2}\o^{-1}=c\_\ell\o^{-1} \eeqno{nitera}
(replacing $\o$), and `acceleration'
\beq  a\equiv ghk=c^2\_\ell k  \eeqno{dofa}
(replacing $k$), to get
\beq\frac{M\bar G}{r_\o^2}=a\cdot{\rm tanh}(a/g). \eeqno{yuio}
We clearly see the close formal analogy with MOND by making the correspondence
\beq g\leftrightarrow \az,~~c\_\ell\leftrightarrow c, ~~h\leftrightarrow \lM,~~
M\bar G\leftrightarrow MG. \eeqno{muop}
Scale invariance holds for the analog of the deep-MOND limit, $a\ll g$ ($\l\gg h$), where eq. (\ref{yuio}) reduces to
\beq \bar M Ggr_\o^{-2}= a^2.  \eeqno{certy}
The analog of the Newtonian limit in MOND is $h\gg \l$, where
\beq  M\bar G/r_\o^2\approx a, \eeqno{zareq}
 which is not scale invariant.
The emergent `interpolating function'\footnote{An interpolating function in MOND is a device used in most existing MOND theories that is put in by hand to interpolate between the standard-dynamics regime of high accelerations and the deep-MOND regime at $a\ll\az$.} for this problem is
$\m(x)={\rm tanh}(x)$, which happens to have just the form and asymptotic behaviours as in MOND.
\par
Some lessons that we can carry to the case of MOND, from this system, whose effective dynamics emerge from general hydrodynamics (itself emergent from microscopics) are as follow:
1. How different constants can enter in various combinations the emergent (approximate) dynamics of the system as new `fundamental' constants. 2. How an interpolating function can emerge. 3. The possible importance of the extra dimensions. The physics of our system is two-dimensional, as it concerns surface waves. However, its dynamics is dictated by, and emerges from what happens in the three-dimensional hydrodynamical system; in particular, communication through the third dimension at the bulk speed of sound. 4. The possibility of fast propagation through the extra dimensions at speeds much exceeding the characteristic speeds of our emergent dynamics. In our case the bulk speed is much higher than that of surface waves.
\par
These points may be relevant to brane-world picture origins of MOND.

\par
And now to MOND itself.
The cosmological constant, $\Lam$, enters local dynamics even in GR (and its nonrelativistic limit) where it adds a (mass independent) acceleration term that is linear in the distance $a=c^2\Lam r/3=c^2 r/\dsr^2=r/\tau^2_0\sim \az r/\dsr$. So, in effect it enters here as a time constant, $\tau_0\equiv c(\Lambda/3)^{-1/2}$, which is of the order of the Hubble time.
\par
But, in the context of MOND we need to understand how $\Lam$ may have entered as an acceleration.\footnote{In this discussion of `boxed' systems I concentrate on the $\az-\Lam$ connection, as most of the suggestions in this vein for an $\az-\aU$ connection revolve on $\Lam$.}
\par
\begin{figure}
\begin{center}
\includegraphics[width=0.6\columnwidth]{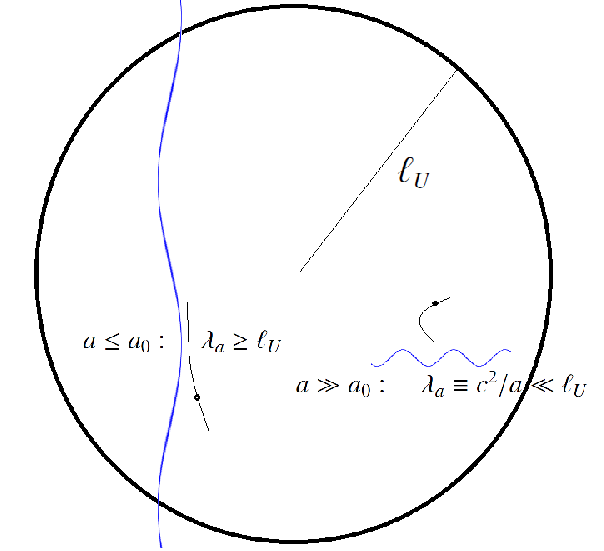}
\end{center}
\caption{Lengths (e.g., Unruh wavelengths) associated with accelerations larger and smaller compared with $\az$, as compared with $\lU\sim\lM$.}\label{unruh}
\end{figure}

This is made plausible by noting that we can associate with an acceleration $a$ a length
$\la\equiv c^2/a$ with crucial significance in various contexts. For example, this is the size of the near field of a radiating accelerating charge,\footnote{So, we may expect that in a presence of dielectric or conducting boundaries a distance $\ell$ from the charge, the electromagnetic field would change character depending on whether $\la\ll \ell$ and the opposite limit.} but more generally, it is the distance to the Rindler horizon, and the characteristic wavelength of the Unruh spectrum associated with a constant-acceleration system.
It may be said then that a system with acceleration $a$ probes to a distance $\la$. This is analogous to the quantum situation, where in line with the uncertainty relation, a system of momentum $P$ probes to distances down to $\sim\hbar/P$.
And, since
\beq a\lessgtr\az ~~~~~ \Leftrightarrow ~~~~~   \lU\lessgtr \ell\_a,  \eeqno{juta}
we can see that systems with $a\gg\az$ probe in this sense to distances much smaller than $\lU$, and are thus not aware of the presence of the cosmological `box'. In contradistinction,
systems with $a\lesssim\az$ do know about the presence of the box and could have different dynamics {\it if one could actually establish a dynamical effect of the box on local physics} (see Fig. \ref{unruh}).
\par
This observation was the basis for the suggestion in Ref. \cite{milgrom99} that the quantum vacuum in our near-de-Sitter Universe provides an inertial frame that is responsible for inertia, in which, indeed, the Unruh-Gibbons-Hawking spectrum changes character between the two acceleration regions below and above $\alam$, in a way that is compatible with MOND.
It was suggested that this kinematic change of character may underlie the dynamical change of character that MOND calls for.
The same observation served as basis for numerous subsequent works; e.g., Refs. \cite{kk11,pa12,verlinde17,ms16,smolin17}.

\subsection{MOND in a brane universe}
MOND, and, in particular, an $\az-\aU$ relation such as (\ref{coinc}), may also result in scenarios where the Universe is a sphere-like submanifold, also called brane, of radius $\lU$ embedded in a higher dimension space-time \cite{milgrom19a}. This brane universe is subject in the embedding space-time to its own dynamics, which, in turn, enters the effective, emergent dynamics that govern physics confined to the submanifold, which is the dynamics we observe (e.g., as MOND).
\par
Such an underlying picture for MOND would be but one example of many suggestions that go under the name of brane-world scenarios.
The last several decades have seen a growth of interest in such pictures of the Universe, which have been constructed to address various issues in particle physics, cosmology, or in gravity. They come in a large variety as regards the geometries and topologies of the embedding space and of the brane itself (including possibly several or many branes). These also differ by the kinds of influences and interactions the brane is subject to. All these attributes are essentially put in by fiat in order to achieve the desired ends (see Refs. \cite{brax04,ms10} for reviews).
\par
The inspiration for pursuing this approach to MOND springs from the fact that if the `dark energy' is in fact a `cosmological constant', as it seems to be, then our Universe has had in recent cosmological epochs a nearly de Sitter geometry.
\par
A (four-dimensional) de Sitter space-time, $dS^4$, can be viewed as a pseudosphere (i.e., one having a metric whose signature is Minkowskian) that is embedded in a five-dimensional Minkowski space-time, $M^5$. If the coordinates in the latter
are $X\!A$, $A=0,1,...,4$, then, the collection of all points satisfying
\beq X^2\equiv X\!A\eta\_{AB}X\!B=\dsr^2, \eeqno{iias}
with the induced metric, is a $dS^4$ of radius $\dsr$ [$\eta\_{AB}=diag(-1,1,1,1,1)$, and in this section I take $c=1$].
\par
For a world line $X\!A(\tau)$ in the $dS^4$, take the $\tau$ derivative\footnote{$\tau$ is the proper time in both the $M^5$ and the $dS^4$.} of eq. (\ref{iias}) to get
 \beq X\!A\eta\_{AB}\dot X\!B=0.  \eeqno{masa}
Thus, $n\!A=X\!A/|X|=X\!A/\dsr$ -- which is the unit radial vector in the $M^5$ at $X\!A$ -- is perpendicular to all tangent vectors to the $dS^4$ at $X\!A$; i.e., it is the local unit normal to the $dS^4$.
\par
Differentiate eq. (\ref{masa}) we get
\beq\dot X\!A\eta\_{AB}\dot X\!B+X\!A\eta\_{AB}\ddot X\!B=0.  \eeqno{masba}
Since $\dot X\!A\eta\_{AB}\dot X\!B=-1$,  we get
\beq  n\!A\eta\_{AB}\ddot X\!B=1/\dsr=\alam. \eeqno{mufa}
Thus, for any world line in the $dS^4$, when viewed as a world line in the $M^5$, its  $M^5$ acceleration, $a\_5\!A=\ddot X\!A$, has a component in the radial direction that is always the constant $\alam$.
\par
The covariant acceleration in the $dS^4$ is
$ a^\m=D^2x^\m/D\tau^2=\ddot x^\m+\Gamma^{\m}_{\n\a}\dot x^\n\dot x^\a$,
with $\Gamma^{\m}_{\n\a}$ the connection in the $dS^4$.
It is then easy to see that the magnitudes of the accelerations in the $M^5$ ($a_5^2=\ddot X\!A\eta\_{AB}\ddot X\!B$) and  that in the $dS^4$ ($a^2=\gmn a^\m a^\n$, where $\gmn$ is the metric in the $dS^4$) are related by
 \beq a\_5^2=a^2+\alam\!2. \eeqno{nuta}
\par
This picture of the $dS^4$ space-time as a sphere of radius $\dsr$, embedded in a flat, one-dimension-up space-time, and with a constant acceleration $\alam=c^2/\dsr$ acting everywhere perpendicular to the $dS^4$, (which is also the radial direction from the center of the embedded sphere) is usually taken only as a mathematical description with no dynamical bearing.
\par
However, in our context, this description of $dS^4$ has inspired the picture in which our universe is a physical, (almost) spherical brane embedded in a one-up-dimensional flat space, and is acted on by a radial force under which is has attained a spherical shape at some equilibrium radius, where the radial force is balanced by the tension inherent in the brane.
\par
A heuristic brane-world picture for MOND based on this has been discussed in detail in Ref. \cite{milgrom19a}, here I outline it very succinctly.
\par
In this simplistic model, I considered a nonrelativistic picture, in which the brane has the topology of a three-dimensional sphere, embedded in a four-dimensional Euclidean space where it is subject to some radial potential  $\fm(r)$ that couples to the mass density of the brane, $\s$, as $\fm\s$. Without further disturbances, the brane attains a spherical shape, centered at the origin of the central force, of such a radius, $\ell\_0$,  where the external force, which per unit volume is $\fm'(\ell\_0)\s$ is balanced by the brane's tension, $T$ ($\s$ and $T$ may depend on the branes radius, and here are taken at the equilibrium radius).
\par
\begin{figure}[h]
	\centering
\includegraphics[width=0.8\columnwidth]{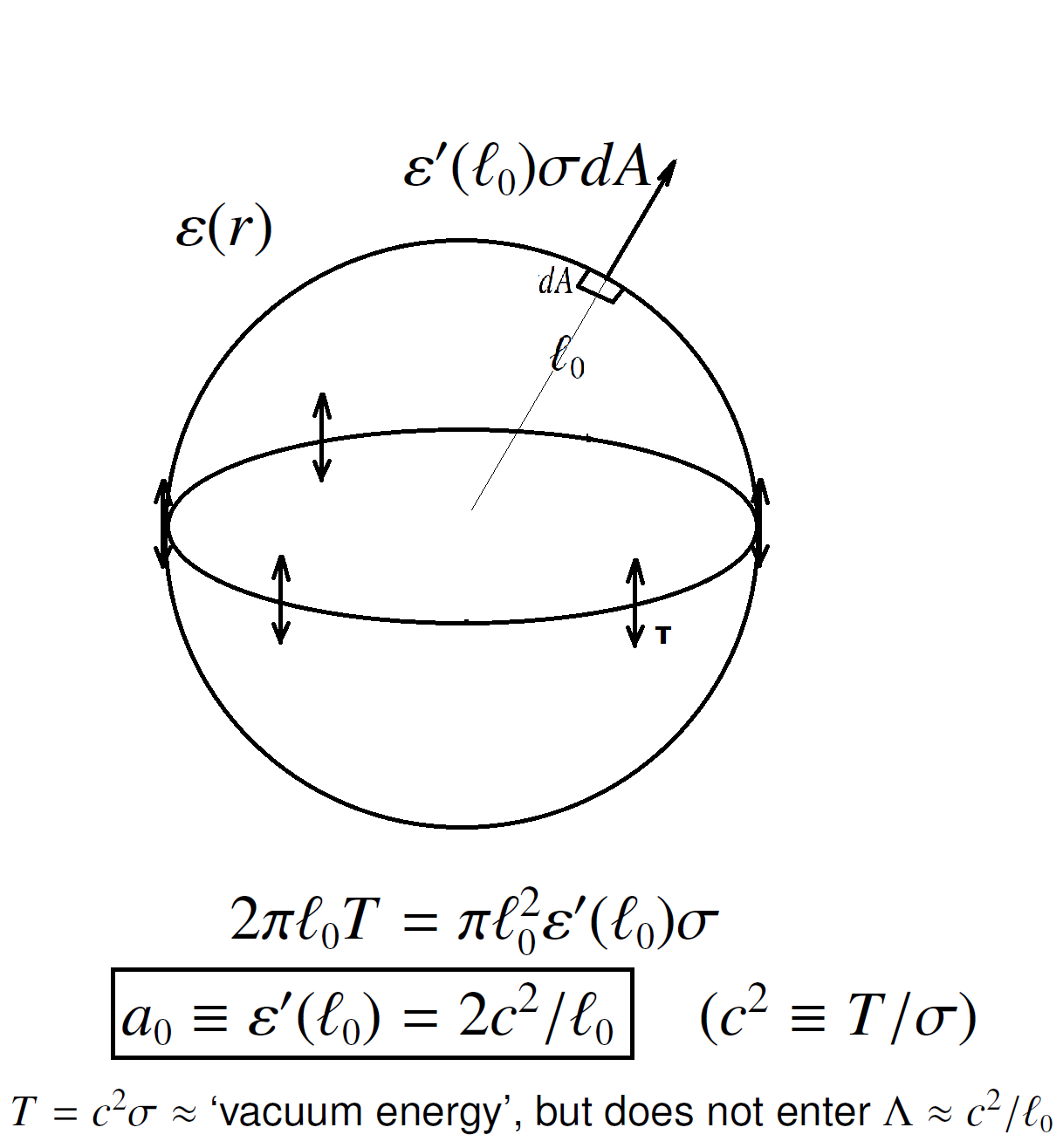}
\caption{Equilibrium of a spherical, two-dimensional brane.}		\label{emptybrane}
\end{figure}
\begin{figure}[h]
	\centering
\includegraphics[width=0.9\columnwidth]{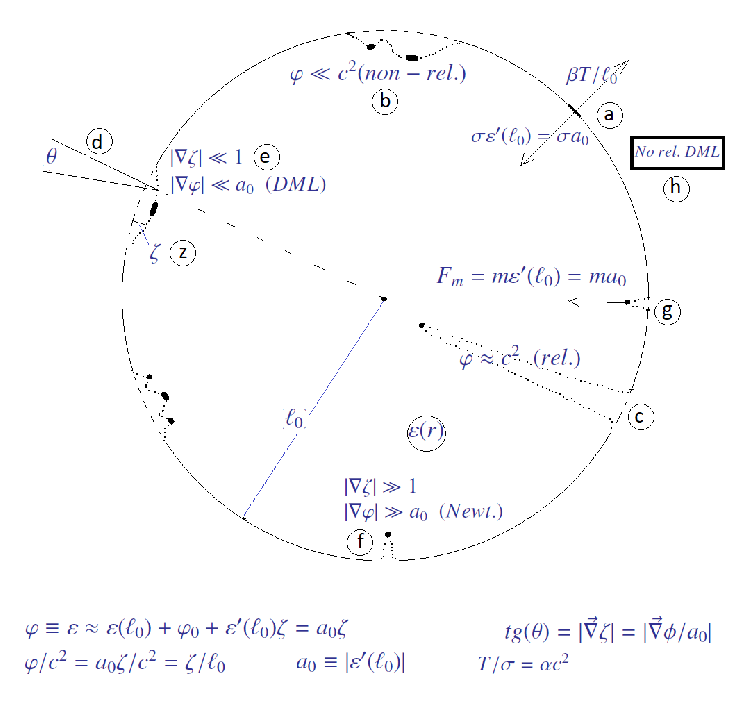}
\caption{Schematics of the brane dynamics: The brane (of density $\s$) and `masses' ($m$)  on it couple to a spherical potential $\fm$ in the embedding space (encircled $\fm$ in the Figure). The brane thus attains an on-average spherical shape of radius $\lz$, for which the force due to brane pressure (tension), $T$, which per unit volume is  $\b T/\lz$ ($\b$ is a geometrical factor of order unity) is balanced by the force per unit volume $\s\fm'(\lz)\equiv\s\az$ (encircled ``a'' in the Figure).
The speed of propagation of perturbations on the brane is $c^2=\a^{-1}T/\s=(\b\a)^{-1}\lz\az$ ($\a$ of order unity).
The nonrelaivistic gravitational potential $\f\equiv \az\z$, where $\z$ is the local departure from sphericity (encircled ``z'').
A shallow depression $\z\ll\lz$, which is equivalent to $\f\ll c^2$, characterizes nonrelativistic gravity (encircled ``b''), an approximation that is broken  where $\z\not\ll\lz$ (encircled ``c''). $\t$ is the angle between the normal to the brane and the radial direction (encircled ``d''). The limit $|\tan\t|\ll 1$ -- the same as $|\grad\z|\ll 1$, and the same as $|\gf|\ll\az$ -- corresponds to the deep-MOND limit (encircled ``e''), while $|\tan\t|\gg 1$ is the Newtonian limit (encircled ``f'').
The $\fm$ force on a mass $m$ is $m\fm'$, which for `nonrelativistic' depressions is $\approx m\fm'(\lz)\approx m\az$ (encircled ``g''). This breaks down when $\f\not\ll c^2$. A depression that is both relativistic ($\z\not \ll 1$), and in the deep-MOND limit ($|\grad\z|\ll 1$) is not possible, as it would have to have an extent $\gg \lz$ (encircled ``h'').}			\label{brane}
\end{figure}

Such a balance in the general case of an $n$-dimensional sphere in an $(n+1)$-dimensional Euclidean space requires that
\beq V_n \ell\_0^n \fm'(\ell\_0)\s=T S_{n-1}\ell\_0^{n-1},   \eeqno{sphapha}
where $V_n \ell\_0^n$ is the `area' of the brane's equatorial plane inside the equator, and $S_{n-1}\ell\_0^{n-1}$ is the `circumference' of the equator (see Fig. \ref{emptybrane} for a depiction of the $n=2$ case). It is easily seen that $S_{n-1}/V_n =n$, and remembering that the `sound speed' on the brane is $c=(T/\s)^{1/2}$,
we get
\beq \az\equiv \fm'(\ell\_0)=n\frac{c^2}{\ell\_0}.  \eeqno{sufata}
The rest of the discussion in Ref. \cite{milgrom19a} includes showing how, indeed, this $\az$, which is an emergent acceleration constant in the brane picture, can also play the role of $\az$, in an emergent MOND dynamics of `masses' that are confined to the brane, and that are equally subject to the potential $\fm(r)$. Such masses  cause indentations or dimples in the brane, where the extra force on the mass itself is balanced by some elastic properties of the brane. This is similar to the famous toy model for gravity by considering a stretched membrane in a constant acceleration field on which masses lie and cause dimple (see Fig. \ref{brane} for more detail).
\par
The spherical geometry of the brane is interpreted by its inhabitants -- who measure it through observations of cosmology -- as a cosmological constant $\Lam=3\ell^2_0$.
If we then identify the speed of small perturbations on the brane, $c$, with the speed of light in the real Universe, we get the desired relation (\ref{coinc}) between $\az$, $c$, and $\Lam$.
\par
The departure of the brane from sphericity, assumed small compared with $\ell\_0$, are interpreted as the gravitational potential in the brane. Other details are to be found in Ref. \cite{milgrom19a} and in Fig. \ref{brane} taken therefrom.
\section{Summary\label{summary}}
There is a numerical relation -- a `coincidence' -- connecting the MOND acceleration constant, $\az$ (as measured from its many appearances in local, galactic dynamics), the speed of light $c$, and characteristic cosmological parameters. The last can be either the Hubble-Lema\^{i}tre constant, $H_0$, or the cosmological constant, $\Lam$.
The relation can also be written as one between $\az$, $c$, and a cosmological
scale length, $\lU$, which can be either the Hubble distance, $c/H_0$, or the `de Sitter radius' $\dsr=(\Lam/3)^{-1/2}$.
In the latter terms, the relation can be written as
\beq \az\sim c^2/\lU.  \eeqno{relsum}
I pointed to several interesting phenomenological corollaries of the numerical coincidence itself, which apply irrespective of possible underlying origins of this relation.
\par
More interestingly, I argue that this relation does bespeak an origin of MOND phenomenology in a deeper theory, where cosmology and local MOND will be understood as two aspects of the same construct, and from which the above relation will follow naturally.
\par
My arguments for this eventuality are based on the following:  On one hand, on the fact that there are already ideas for explaining this relation in underlying schemes for MOND. And, on the other hand, on the fact that similar relations appear in other physics contexts, where they indeed show up in effective theories that emerge, under restricted circumstances, as approximations of more fundamental theories.


\begin{thebibliography}{}
\bibitem{milgrom83a}M. Milgrom  (1983). A modification of the Newtonian dynamics as a possible alternative to the hidden mass hypothesis. Astrophys. J. 270, 365.
\bibitem{milgrom83b}M. Milgrom  (1983). A modification of the Newtonian dynamics - Implications for galaxies. Astrophys. J. 270, 371.
\bibitem{milgrom83c}M. Milgrom  (1983). A modification of the newtonian dynamics : implications for galaxy systems. Astrophys. J. 270, 384.
\bibitem{fm12}B. Famaey \& S. McGaugh (2014). Modified Newtonian Dynamics (MOND): Observational Phenomenology and Relativistic Extensions. Liv. Rev. Rel. 15, 10 (2012).
\bibitem{milgrom14c}M. Milgrom (2014). The MOND paradigm of modified dynamics.  Scholarpedia 9(6), 31410 (continually updated).
\bibitem{milgrom14a}M. Milgrom (2014). MOND laws of galactic dynamics. Mon. Not. R. Astron. Soc. 437, 2531.
\bibitem{milgrom89}M. Milgrom (1989). Alternatives to Dark Matter. Comm. Astrophys. 13 (4), 215.
\bibitem{milgrom94}M. Milgrom (1994). Dynamics with a Nonstandard Inertia-Acceleration Relation: An Alternative to Dark Matter in Galactic Systems. Ann. Phys. 229, 384.
\bibitem{efstathiou90}G. Efstathiou, W.J. Sutherland, \& S.J. Maddox (1990). The cosmological constant and cold dark matter. Nature 348, 705.
\bibitem{milgrom99}M. Milgrom (1999). The modified dynamics as a vacuum effect. Phys. Lett. A   253,  273.
\bibitem{li18}P. Li et al. (2018). Fitting the radial acceleration relation to individual SPARC galaxies. Astron. \& Astrophys. 615, A3
\bibitem{milgrom88}M. Milgrom (1988). On the Use of Galaxy Rotation Curves to Test the Modified Dynamics. Astrophysical J. 333, 689.
\bibitem{milgrom14b}M. Milgrom (2014). Gravitational waves in bimetric MOND.
    Phys. Rev. D 89, 024027.
\bibitem{milgrom11}M. Milgrom (2011). Gravitational Cherenkov Losses in Theories Based on Modified Newtonian Dynamics.
    Phys. Rev. Lett. 106, 111101.
\bibitem{cl17}P.M. Chesler \& A. Loeb (2017). Constraining Relativistic Generalizations of Modified Newtonian Dynamics with Gravitational Waves. Phys. Rev. Lett. 119, 031102.

\bibitem{zlosnik07}T.G. Zlosnik, P.G. Ferreira, and G.D. Starkman (2007). Modifying gravity with the aether: An alternative to dark matter. Phys. Rev. D 75, 044017
\bibitem{milgrom09b}M. Milgrom (2009). Bimetric MOND gravity. Phys. Rev. D 80, 123536
\bibitem{blh17}L. Blanchet and L. Heisenberg (2017). Dipolar dark matter as an effective field theory. Phys. Rev. D. 96, 083512
\bibitem{milgrom19}M. Milgrom (2019). Noncovariance at low accelerations as a route to MOND. Phys. Rev. D 100, 084039
\bibitem{milgrom15}M. Milgrom (2015). Road to MOND: A novel perspective. Phys. Rev. D 92, 044014
\bibitem{kk11}F.R. Klinkhamer and M. Kopp (2011). Entropic Gravity, Minimum Temperature, and Modified Newtonian Dynamics. Mod. Phys. Lett. A 26, 2783
\bibitem{pa12}E. Pazy and N. Argaman (2012). Quantum particle statistics on the holographic screen leads to modified Newtonian dynamics. Phys. Rev. D 85, 104021
\bibitem{verlinde17}E. Verlinde (2017). Emergent Gravity and the Dark Universe. SciPost Phys. 2, 016
\bibitem{ms16}M. Milgrom \& R.H. Sanders (2016). Perspective on MOND emergence from Verlinde's "emergent gravity" and its recent test by weak lensing. arXiv:1612.09582
\bibitem{smolin17}L. Smolin (2017). MOND as a regime of quantum gravity. Phys. Rev. D 96, 083523
    \bibitem{milgrom19a}M. Milgrom (2019). MOND from a brane-world picture. arXiv:1804.05840; in `Jacob Bekenstein, the conservative revolutionary. Eds. L. Brinks et al., World Scientific
\bibitem{brax04}P. Brax, et al. (2004).  Brane World Cosmology.  Rept. Prog. Phys. 67, 2183
\bibitem{ms10}R. Maartens and K. Koyama (2010). Brane-World Gravity. Liv. Rev. Rel. 13, 5

\end{thebibliography}
\end{document}